# Static Skew Compensation in Multi Core Radio over Fiber systems for 5G Mmwave Beamforming

Thomas Nikas [1], Evangelos Pikasis [2], Dimitris Syvridis [1]
(1): Dept. of Informatics and Telecommunications National and Kapodistrian University of Athens, tnikas@di.uoa.gr
(2): Eulambia Advanced technologies Ltd, evangelos.pikasis@eulambia.com

*Abstract*—Multicore fibers can be used for Radio over Fiber transmission of mmwave signals for phased array antennas in 5G networks. The inter-core skew of these fibers distort the radiation pattern. We propose an efficient method to compensate the differential delays, without full equalization of the transmission path lengths, reducing the power loss and complexity.

## I. INTRODUCTION

The centralization of the Baseband Units (BBUs) and simplification of the remote radio heads (RRHs) are key elements for the forthcoming 5G networks. The bandwidth requirements push the operating frequencies to mmwave and benefits from using dense networks of pico and fempto-cells [1]. Radio over Fiber (RoF) becomes a promising technique to serve network densification and flexible resource allocation, greatly simplifying the RRHs. The Radio Frequency (RF) chain at RRH amplifies and feeds the antenna, surpassing the electrical LO and frequency mixing stages.

Antenna beam-forming (BF) is utilized to alleviate the free space path loss and through wall attenuation, important factors for efficient radio coverage at mmwave frequencies. The extended bandwidth and narrow beams required for throughput maximization imposes the use of true time delay (TTD) elements for phase shifting of the currents feeding the antenna to avoid beam squint. Efficient TTD optical beam forming networks (OBFNs) have been proposed which provide multiple squint-free RF beams [2]. Since the OBFNs require thermal stabilization and multiple control signals, it is more beneficial to locate them at the CU. The modulated optical carriers undergo the required time delays per antenna element and then transmitted to the RRHs. The detected mmwave signals are amplified and feed the antenna elements. In order to preserve the RF phase relations, equal optical and electrical path lengths are required after the OBFN. The electrical path length equality can be easily achieved since the RF circuitry at RRH consists of photodiodes, transimpedance and mmwave amplifiers. Most challenging is the optical path length equality, taking into account that the distance between the CU and the RRHs can range from several hundred meters to some kilometers. A tolerant to environmental variations solution is using multi-core fibers (MCFs) for RoF transmission instead of an SMF bundle. MCFs are less prone to effective length variations due to temperature and mechanical stress perturbations. Nevertheless, homogenous MCFs are reported to exhibit inter-core maximum static skew in the order of 0.5 ns/km, while the temperature imposed maximum dynamic skew is 0.06 ps/km/°C [3]. It is obvious from these figures that the major factor affecting the antenna radiation pattern is the value of the static inter-core skew. The compensation of the static skew can be performed either in the optical or the electrical domain. Switchable optical delay lines can be inserted at the output of the OBFN introducing additional power loss.

In this paper, we perform theoretical analysis and simulations quantifying the MCF static skew effects on the antenna array factor (AF) and RF channel frequency response and introduce an efficient compensation process, without full equalization of the transmission path lengths.

## II. ANALYTICAL MODEL

In order to evaluate the MCF skew effects on the antenna AF we consider $N$ $\lambda/2$ spaced isotropic radiating elements in a uniform linear array (ULA). Each element exhibits flat frequency response within the signal bandwidth. The modulation format proposed for 5G is the orthogonal frequency division modulation (OFDM), so the baseband data signal s(t) is expressed as $s(t) = \sum_{k=-\frac{N_F}{2}}^{\frac{N_F}{2}} X_k e^{\frac{j2\pi kt}{T_s}}$ where $X_k$ is the complex symbol modulating the $k$-th subcarrier, $T_s$ is the OFDM symbol duration without the cyclic prefix and $N_F$ is the IFFT length. The baseband signal is up-converted to mmwave. After optical modulation and passing through the OBFN, the optical signal outputs are launched to N cores of an MCF, each core corresponding to an antenna element. After transmission, the inter-core skew imposes differential time delays $\tau_i$, $i=1,...,N$ between each core and the core with minimum effective path length, which is considered as reference (zero delay). At RRH, the $N$ photocurrents are read as $I(i,t) = I_{0,i} \cdot s(t-\tau_i) \cdot e^{j2\pi f_c t}$, $i = 1,..., N$ where $I_{0,i}$ is the normalized complex amplitude of the *i*-th photocurrent, $I_{0,i} = |I_{0,i}|e^{-j(i-1)\pi\sin\theta_0}$ and $|I_{0,i}|$ is the same for all *i*, if no side-lobe suppression measures are considered. The term $\pi\sin\theta_0$ is the phase delay step imposed by the OBFN to steer the beam to $\theta_0$. The photocurrents feed the isotropic antenna elements and since the electrical far field $E_R(i,t)$ of the transmitted wave is proportional to $I(i,t)$, the received at distance $r_i$ field from the *i*-th antenna element is $E_R(i,t) = aI(i,t)e^{-j2\pi f_c r_i/c}$, where *a* symbolizes the linear electrical field – antenna current relation and line of sight path loss. The factor *a* is assumed time invariant and frequency independent within the signal bandwidth, in order to analyze solely the effects of inter-core skew on the AF.

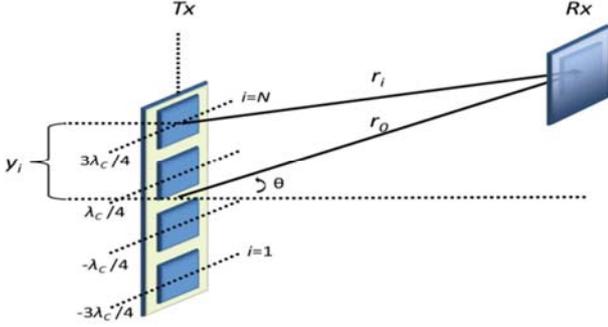

Fig. 1: The ray tracing model for N=4, $\lambda_c = \frac{c}{f_c}$

The receiver antenna is located at distance $r_0$ from the transmitter and considered isotropic (fig.1). The distance $r_0$ is set to several thousands of free space wavelengths $\lambda_c$, ensuring that the receiver is located at the transmitter far field. The distance of each transmitter antenna element to the receiver is

$$r_i = \sqrt{r_0^2 + y_i^2 - 2 \cdot r_0 \cdot y_i \cdot \cos(\tfrac{\pi}{2} - \theta)}, -\tfrac{\pi}{2} \le \theta \le \tfrac{\pi}{2} \quad (1a)$$

$$y_i = \left(2\left(i - \tfrac{N}{2}\right) - 1\right)\tfrac{\lambda_c}{4}, \qquad i = 1, \dots, N \quad (1b)$$

The total field $E_R(t)$ at the receiver antenna is the vector sum of all the fields generated from each transmitter antenna element

$$E_R(t) = \sum_{i=1}^{N} aI_{0,i}s(t - \tau_i)e^{j2\pi f_c t} e^{-j2\pi f_c r_i/c} \quad (2)$$

The receiver antenna current can be read as $b \cdot E_R(t)$. The factor *b* symbolizes the electrical field – antenna current relation, assumed to have the same time invariability and frequency independence properties of *a*. After down-conversion from $f_c$, analog to digital conversion and applying FFT, the complex baseband received signal $R(t)$ is transformed to the discrete frequency domain:

$$R_k = ab \sum_{i=1}^{N} I_{0,i}X_k e^{-j2\pi(f_c+k/T_s)\tau_i} e^{-j2\pi f_c r_i/c} \quad (3)$$

The inter-core skew delays $\tau_i$ impose phase delays proportional to the subcarrier frequency. If we consider the transmitted OFDM symbols as pilot symbols for channel estimation, then a zero forcing frequency domain equalizer will output the transfer function:

$$H_k = \frac{R_k}{X_k} = ab \sum_{i=1}^{N} I_{0,i}e^{-j2\pi(f_c+k/T_s)\tau_i} e^{-j2\pi f_c r_i/c} \quad (4)$$

In (4), the transmitted QAM symbols $X_k$ are perfectly demodulated at the receiver in the absence of noise. Since $r_i$ is a function of the angle $\theta$, the quantity

$$P(\theta) = \sum_{k=-\frac{N_F}{2}}^{\frac{N_F}{2}} \frac{|H_k|^2}{N_{Fu}} \quad (5)$$

is the total received power at $\theta$, normalized to the number of the used subcarriers $N_{Fu}$, with $N_{Fu} \le N_F$. Using (4) and (5) we can calculate the frequency response and the normalized total received power at any $\theta$ and consequently the array factor $P(\theta)$.

Since the analytical model does not encounter the effects of noise, the parameters $a, b$ are only scaling factors and can be set to $ab = 1$. In all cases, the total transmitted electrical power is normalized to

$$\sum_{i=1}^{N} |I_{0,i}|^2 = 1 \quad (6)$$

### III. STATIC SKEW PARADIGM AND COMPENSATION

At first, we calculate the transfer function $|H_k|^2$ and the AF $P(\theta)$ for a ULA of *N*=4 elements, at a carrier frequency of $f_c = 26$ GHz. The IFFT length is set to 4096 and the OFDM signal is composed of 3332 subcarriers with 240 kHz spacing, $T_s = 1/(240 \text{ kHz})$, resulting in a signal bandwidth of 800 MHz. It should be noted that our analytic model is more general in terms of modulation formats and not specifically based on OFDM. On the other hand, OFDM is straightforward in acquiring the channel transfer function in frequency domain. The antenna elements are fed with uniform current distribution and $|I_{0,i}| = 1/\sqrt{N}$ following the total power constraint in (6). The beam steering angle was set to $\theta_0=0°$ and the inter-core differential delays set to zero. The corresponding $|H_k|^2$ at $\theta_0$ and $P(\theta)$ values are depicted in fig.2 and fig.3 (blue lines). $|H_k|^2$ is flat over the signal bandwidth and the AF corresponds to a typical ULA of four $\lambda/2$ spaced elements.

Next, we calculate $|H_k|^2$ at $\theta_0$ and $P(\theta)$ imposing delays with values of 0.00, 0.44, 0.50, 0.10 ns which correspond to 1 km of RoF transmission through cores 2, 3, 6, and 7 of the MCF studied in [3]. Core 2 is considered as reference. $|H_k|^2$ and $P(\theta)$ are depicted in fig.2 and fig.3 with red lines. As expected, $|H_k|^2$ resembles to the transfer function of a frequency selective channel. This is attributed to the vector summing of fields produced from different antenna elements with variable delays, as in the case of multipath propagation. Moreover, the antenna radiation pattern is severely distorted due to the phase misalignment of the RF signals feeding the antenna elements, originated from the differential delays introduced through the RoF transmission.

The obvious compensation method of the inter-core skew effects is to add optical or electrical delay lines in order to perfectly equalize the end to end delays, as stated in the introduction. Nevertheless, the severe radiation pattern distortion can be partially compensated, if we apply phase leads $\varphi_i$ to the antenna elements equal to the residuals modulo $2\pi$ of the differential phase lags introduced by the static inter-core skew at the carrier frequency $f_c$, $\varphi_i = \text{rem}(2\pi f_c \tau_i, 2\pi)$. The compensated transfer function $\widetilde{H_\kappa}$ is:

$$\widetilde{H_\kappa} = \sum_{i=1}^{N} I_{0,i} e^{-j2\pi(f_c + k/T_s)\tau_i} e^{-j2\pi f_c r_i/c} e^{j\varphi_i} \quad (7)$$

The calculated $|\widetilde{H_\kappa}|^2$ at $\theta_0$ and $P(\theta)$ using (7) are depicted in fig.2 and fig.3 (black lines). In fig. 3, the array factor shape is restored, with slightly stronger side-lobes and shallower nulls. In fig.2, $|H_k|^2$ shows a frequency selective behavior as expected.

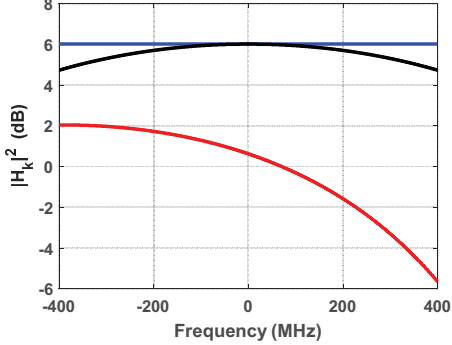

Fig. 2: Calculated frequency response. Blue: w/o skew, red: w/ skew, black: w/ compensation

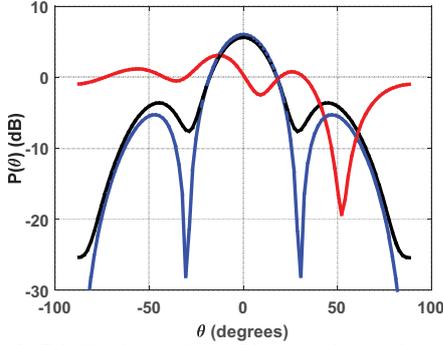

Fig. 3: Calculated array factor. Blue: w/o skew, red: w/ skew, black: w/ compensation

The phase term $2\pi(f_c + k/T_s)\tau_i$ in (7) can be expressed as $2m\pi + \text{rem}(2\pi f_c \tau_i, 2\pi) + \frac{2\pi k}{T_s}\tau_i$, and so we get

$$\widetilde{H_\kappa} = \sum_{i=1}^{N} I_{0,i} e^{-j2\pi(k/T_s)\tau_i} e^{-j2\pi f_c r_i/c} \quad (8)$$

Recalling that $I_{0,i} = |I_{0,i}| e^{-j(i-1)\pi \sin \theta_0}$ and taking into account that the phase lags imposed on the antenna currents compensate the phase offsets due to the path length differences of $r_i$ in the direction $\theta_0$, (8) can be expressed as

$$\widetilde{H_\kappa} = e^{-j2\pi f_c r_0/c} \sum_{i=1}^{N} |I_{0,i}| e^{-j2\pi(k/T_s)\tau_i} \quad (9)$$

The term $e^{-j2\pi f_c r_0/c}$ imposes a constant phase shift, not affecting the channel frequency response and can be neglected. So (9) is reduced to $\widetilde{H_\kappa} = \frac{1}{\sqrt{N}} \sum_{i=1}^{N} e^{-j2\pi(k/T_s)\tau_i}$ for uniform current distribution of the antenna elements.

In order to evaluate the inter-core skew effects on the wireless channel, we calculated the coherence bandwidth $B_c$. The definition of $B_c$ is based on the complex autocorrelation function $R_q = \sum_{k=-\infty}^{+\infty} \widetilde{H_k} \widetilde{H_{k-q}^*}$ of the frequency response $\widetilde{H_\kappa}$ [4]. $B_c^{0.9}$, $B_c^{0.5}$ are the values of $\frac{q}{T_s}$ where $|R_q|$ has decreased to 0.9 or 0.5 of its maximum value respectively. Setting the summing limits of $k$ to $\pm N_{Fu}/2$ and after algebraic manipulation and approximations we get:

$$R_q = \frac{1 + (N-1)(N_{Fu}+1)}{N} \sum_{i=1}^{N} e^{-j2\pi\left(\frac{q}{T_s}\right)\tau_i} \quad (10)$$

Next we calculate $|R_q|$ using (10), normalize the results to $|R_{q,n}| = |R_q|/|R_0|$ and depict them in fig.4:

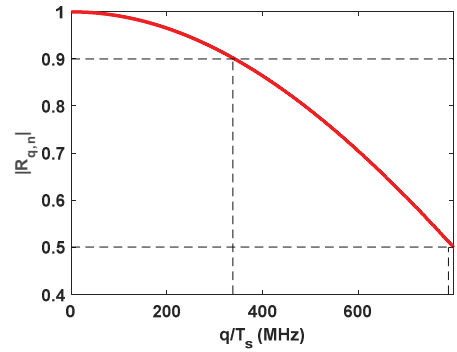

Fig. 4: The autocorrelation function $|R_{q,n}|$ using eq. 10. Coherence bandwidth estimation: $B_c^{0.9} \cong 338\ MHz$, $B_c^{0.5} \cong 789\ MHz$.

The $B_c^{0.5}$ value is fairly equal to the signal bandwidth in this particular case, by only compensating the residual $\varphi_i$ and most of the static skew can be left uncompensated. In any case, if the uncompensated static skew limits $B_c$ to values lower than the signal bandwidth, a partial compensation using electrical or optical means can be applied to restore the coherence bandwidth.

ACKNOWLEDGMENT

This work was supported in part by the EU project blueSPACE (H2020-ICT2016-2, 762055).